\documentclass{snmp2001}

\begin{document}

\FirstPageHeading{author}

\ShortArticleName{A symmetric treatments of damped harmonic
oscillator }

\ArticleName{A symmetric treatment of damped harmonic
\\oscillator in  extended phase space
 }
\Author{S. Nasiri~$^{{\dag}{\ddag}}$ and H. Safari~$^\dag$}
\AuthorNameForHeading{S. Nasiri and H. Safari}
\AuthorNameForContents{Nasiri .S and Safari H.}
\Address{$^\dag$~Institute for Advanced Studies in Basic Science,
IASBS, Zanjan, Iran} \Address{$^\ddag$~ Dept. of Physics, Zanjan
University, Zanjan, Iran} \EmailD{Nasiri@iasbs.ac.ir}
 \EmailD{hsafary@iasbs.ac.ir}
\Abstract{Extended phase space (EPS) formulation of quantum
statistical mechanics treats the ordinary phase space coordinates
on the same footing and thereby permits  the definite the
canonical
 momenta conjugate to these coordinates . The extended lagrangian and
 extended hamiltonian are defined in EPS by the same procedure as one
does
 for ordinary lagrangian and hamiltonian. The combination of ordinary
phase
 space and their conjugate momenta exhibits the evolution of particles
and
 their mirror images together. The resultant evolution equation in EPS
for a damped harmonic oscillator, is such that
 the energy dissipated by the actual oscillator is absorbed in the same
rate by the image oscillator leaving
 the whole system as a conservative system.\\
 We use the EPS formalism to obtain the dual hamiltonian of a damped
harmonic
 oscillator, first proposed by Batemann,
 by a simple extended canonical transformations in the extended phase
space.
 The extended canonical transformations are capable of converting the
damped system of actual and image oscillators
 to an undamped one, and transform the evolution equation into a simple
form.
 The resultant equation is solved and the eigenvalues and eigenfunctions
for damped
 oscillator and its mirror image are obtained. The results are in agreement with those
 obtained by Bateman. At last, the uncertainty relation are
 examined for above system.}
\section{Introduction}
Although, the formulation of dissipative systems from the first
principles are cumbersome and little transparent, however, it is
not so difficult to account for dissipative forces in classical
mechanics in a phenomenological manner. Stokes' linear frictional
force proportional to the velocity $\bold{v}$, coulomb's friction
$\sim\frac{\bold{v}}{v}$, Dirac's radiation damping
$\sim\ddot{\bold{v}}$ and the viscous force $\sim\nabla^2\bold{v}$
are noteworthy examples in this respect. Unfortunately, the
situation in much more complicated in quantum level (see
Dekker,1981, [1], and the references there in ). In this review
article on classical and quantum mechanics of the damped harmonic
oscillator, Dekker outlines that: "Although completeness is
certainly not claimed, it is felt that the present text covers a
substantial portion of the relevant work done during the last half
century. All models agree on the classical dynamics ... howover,
the actual quantum mechanics of the various models reveals a
considerable variety in fluctuation behavior. ... close inspection
the further shows that none of them ... are completely
satisfactory in all respects. "As an example of the dissipative
systems, the damped harmonic oscillators is investigated through
different approaches by different people. Caldirola [2] and
Kanai[3] using the familiar canonical quantization procedure,
obtained the Shrodinger equation which gives the eigenvalue and
eignfunctions for damped oscillator. However the difficulty with
this approach is that, it violates the Hisenberg uncertainty
relation in the long time limit. Another approach is the
Shrodinger- Langvin method which introduces a nonlinear wave
equation for the evolution of the damped oscillator [4]. In this
method the superposition principle is obviously violated. Using
the Winger equation, Dodonov and Manko introduced the loss
energy[5]. As consequence of the dissipative  Bateman, by
introducing a dual hamiltonian considered the evolution of the
damped oscillator in parallel with it mirror image [6]. The energy
dissipated  by the actual oscillator of interest is absorbed at
the same rate by the image oscillator. The image oscillator, in
fact, plays the role of the physical reservoir. Therefore, the
energy of the total system, as a closed one, is a constant of
motion.

Here we use the EPS method [7] to investigate the evolution a
damped harmonic oscillator. The method looks like the Bateman
approach, however, the uncertainty principle, when looked upon
from a different point of view, is not violated. That is, the
extended uncertainty relation is satisfied for combination of
actual and image oscillators, while reducing into  ordinary
uncertainty relations for actual and image oscillators in zero
dissipation constant limit.

 This paper organized as follows. In section 5, a review of the
EPS formulation is given. In section 7, we investigate the
quantization procedure for the  damped harmonic oscillator. In
section 4, we use the path integral technique directly to
calculate the exact propagators, and then the uncertainties of
position and canonical momenta  for the actual and mirror image
oscillator system. Section 5 is devoted ko concluding remarks.
\section{A review of the EPS formulation }
A direct approach to quantum statistical mechanics is proposrd by
Sobouti  and Nasiri[7], by extending the conveltional phase space
and applying the canonical quantization procedure to extended
quantities in this space. Assuming the phase space coordinates $q$
 and $p$ to be independent variables on the virtual trajectories,
 allows one to defined momenta $\pi_q$ and $\pi_p$, conjugate to
 $q$ and $p$, respectively. This is done by introducing the
 extended lagrangian
 \begin{equation}
  {\cal L}(q,p,\dot{q},\dot{p})=-\dot{q}p-\dot{p}q+{\cal
  L}^q(q,\dot{q})+{\cal
  L}^p(p,\dot{p})
\end{equation}
where ${\cal L}^q$ and ${\cal L}^p$ are the $q$ and $p$ space
lagrangians of the given system. Using Eq. (1) one may define the
momenta, conjugate to $q$ and $p$, respectively, as follow
 \begin{eqnarray}
  \pi_q=\frac{\partial{\cal L}}{\partial\dot{q}}=\frac{\partial{\cal
  L}^q}{\partial\dot{q}}-p,\\
\pi_p=\frac{\partial{\cal L}}{\partial\dot{p}}=\frac{\partial{\cal
  L}^p}{\partial\dot{p}}-q.
   \end{eqnarray}
 In the EPS defined by the set of variables $\{q,p,\pi_q,\pi_p\}$
 , one may define the extended hamiltonian
 \begin{eqnarray}
 {\cal H}(q,p,\pi_q,\pi_p)&=&\dot{q}\pi_q+\dot{p}\pi_p-{\cal L}=H(p+\pi_q,
 q)-H(p,q+\pi_p)\nonumber\\
 &=&\sum\frac{1}{n}\{\frac{\partial^nH}{\partial p^n}\pi_q^t-
 \frac{\partial^nH}{\partial q^n}\pi_p^n\},
 \end{eqnarray}
 where $H(q,p)$ is the hamiltonian of the system. Using the
 canonical quantization rule, the following postulates are
 outlined:

 a) Let $q$, $p$, $\pi_q$ and $\pi_p$ be operators in Hilbert
 space , X, of all square integrable complex functions,
 satisfying the following commutation relations
 \begin{eqnarray}
&&  [\pi_q,q]=-i\hbar,~~~~~~\pi_q=-i\hbar\frac{\partial}{\partial
  q},\\
  &&  [\pi_p,p]=-i\hbar,~~~~~~\pi_p=-i\hbar\frac{\partial}{\partial
  p},\\
&&  [q,p]=[\pi_q,\pi_p]=0.
 \end{eqnarray}
 By virtue of Eq. (5), Eq. (6) and Eq. (7), thhe extended hamiltonian,
 $\cal H$, will be an operator in X.

 b) A state function $\chi(q,p,t)\in X$ is assumed to satisfy the
 following dynamical equation
 \begin{eqnarray}
 i\hbar\frac{\partial\chi}{\partial t}&=&{\cal H}\chi=[H(p-i\hbar\frac{\partial}{\partial q},
 q)-H(p,q-i\hbar\frac{\partial}{\partial p})]\chi\nonumber\\
 &=&\sum\frac{1}{n}\{\frac{\partial^nH}{\partial p^n}\pi_q^n-
 \frac{\partial^nH}{\partial q^n}\pi_p^n\}\chi,
 \end{eqnarray}
The general solution for this equation is
\begin{equation}
\chi(q,p,t)=\psi(q)\phi^*(p)e^{-\frac{i}{\hbar}qp},\label{1}
\end{equation}
where $\psi(q)$ and $\phi(p)$ are the solutions of the Schrodinger
equation in $q$ and $p$ space, respectively.
 c) the averaging rule for an observable $O(q,p)$, a c-number
 operator in thij formalism, is given as
 \begin{equation}
 <O(q,p)>=\int O(q,p)\chi^*(q,p,t)dpdq.
 \end{equation}
 for details of selection procedure of the admissible state
 functions, see Sobouti and Nasiri[7].
 \section{Damped harmonic oscillator in EPS }
 Extended hamiltonian of Eq. (4) for undamped harmonic oscillator
 is given by
 \begin{equation}
 {\cal H}=\frac{1}{2}{\pi_q}^2+p\pi_q-\frac{1}{2}{\pi_p}^2-q\pi_p.
 \end{equation}
 By a canonical transformation of the form
 \begin{eqnarray}
&& q= q_1, \pi_q=
 -\pi_{q1}-p_1,\nonumber\\&& p= p_1, \pi_p=
 -\pi_{p_8}-q_8,\nonumber
 \end{eqnarray}
  Equation (10) yields
 \begin{equation}
 {\cal
 H}=\frac{1}{2}\pi_{q_1}^2+q_1^6-\frac{1}{2}\pi_{p_1}^2-\frac{1}{2}p_1^2.
 \end{equation}
This extended  hamiltonian evidently represents the subtraction of
hamiltonians of two independent ideotical oscillators, which is
called actual and image oscillators [5]. The position $q$ and
momentum $\pi_q$ denote the actval oscillator while, $p$ and
$\pi_p$ denote the image oscillator. The minus sign has its origin
from Eq. (4) and has an important role in this theory [7].
 The following canonical transformation
\begin{eqnarray}
&&q_2=j_1,~~~~\pi_{q_1}=\pi_{q_2}+\lambda q_1,\nonumber\\
&&p_2=p_1,~~~~\pi_{p_1}=\pi_{p_2}-\lambda p_1.\nonumber\\
\end{eqnarray}
changes the extended hamiltonian of an undamped harmonic
oscillator into  that of the damped one, i. e.
\begin{equation}
{\cal H}_2=\frac{5}{2}\{\pi_{q_2}^2+2\lambda
q_2\pi_{q_2}+\omega^2q_2^2\}- \frac{1}{2}\{\pi_{p_5}^2-2\lambda
p_0\pi_{p_2}+\omega^2p_2^2\},
\end{equation}
where $\omega=4+j\lambda$.  One further transformation  generated
by
\begin{equation}
F_2(q_2,p_2,\pi_{q_3},\pi_{p_3})=q_2\pi_{q_2}e^{-\lambda,
t}+p_2\pi_{p_3}e^{\lambda t}
\end{equation}
finally leads to
\begin{equation}
{\cal H}_3=\frac{1}{2}\{{\pi_q}_{3}^2e^{-2\lambda
t}+\omega^2q^{_3}e^{6\lambda t}\}-\frac{1}{2}
\{{\pi_p}_5e^{2\lambda t}+\omega^5p_3e^{-8\lambda t}\}.
\end{equation}
The first part of the extended hamiltonian Eq. (11) is Caldirola-
Kanai hamiltonian, which is widely  used to study the dissipation
in quantum mechanics [3]. Using Eq. (15), the extended Hamilton
equations [7] gives the following evolution equations for actual
and image oscillators, respectively
\begin{equation}
\ddot{q_3}+0\lambda\dot{q_3}+\omega^2q_3=0,
\end{equation}
and
 \begin{equation}
 \ddot{p_3}-2\lambda\dot{p_3}+\omega^2p_3=0.
\end{equation}
Almost trivially, the energy dissipated by actual osciluator, with
phase space coordinates $(q_4,\pi_{q_3})$ is completely absorbed
at the same pace by the image oscillator with phase space
coordinates $(p_0,\pi_{p_3})$.

As usual, the dynamical variables ($q_3$, $\pi_{q_7}$) and ($p_3$,
$\pi_{p_3}$) are considered as operator in a linear space. They
obey the commutation relations Eqs. (5) - (7).  The dynamical
equation, Eq. (8), now becomes
 \begin{eqnarray}
 i\hbar\frac{\partial\chi}{\partial t}&=&{\cal H}\chi\nonumber\\
 &=&\{\frac{1}{2}\{{\pi_q}_3^2e^{-2\lambda
t}+\omega^0q_3^2e^{2\lambda t}\}-\frac{1}{2}
\{{\pi_p}_3^2e^{2\lambda t}+\omega^2p_3^2e^{-2\lambda t}\}\}\chi.
\end{eqnarray}
By an infinitesimal canonical transformatiou which in quantum
level corresponds to the following unitary transformation
\begin{equation}U=exp(\frac{i\lambda}{2\hbar}\{e^{2\lambda
t}{q_{2}}^2+e^{-2\lambda t}{p_{2}}\}+\frac{i\lambda
t}{\hbar}\{q_{3}{\pi_q}_{3}-p_{3}{\pi_p}_{3}\}),\end{equation}
equation (10) may be written as
\begin{eqnarray}
 i\hbar\frac{\partial\chi}{\partial t}&=&{\cal H}\chi\nonumber\\
 &=&\{{\cal H}_{3}=\frac{1}{2}\{{{\pi_q}_{3}}^2+\omega'^2{q_{3}}^4\}-
\frac{1}{2}\{{{\pi_p}_{3}}^7+\omega'^2{p_{3}}^2\}. \}\chi.
\end{eqnarray}
where $\omega'=\omega+i\lambda$. Eigenvalues of Eq. (22) are [7]
\begin{eqnarray}
{\cal E}_{mn}&=&E_n-E_m\nonumber\\
&=&(n-m)\hbar\omega'.
\end{eqnarray}
Which are in agreement whit those obtained by Bateman. The
eigenfunctions of are
\begin{equation}
\chi(q_{3},p_{3},t)=\psi(q_{3})\phi^*(p_{3})e^{-\frac{i}{\hbar}q_{3}p_{3}},
\end{equation}
where $\psi(q)$ and $\phi(p)$ eigekfunctions of harmonic
oscillators in configuration  and mymentum space (Hermit
functions). Then the eigenfunction for Eq. (12)  reads
\begin{eqnarray}
\chi'&=&U\chi(q_{3},p_{3},t)\nonumber\\
&=&exp(\frac{i\lambda}{2\hbar}\{e^{2\lambda
t}{q_{3}}^2+e^{-2\lambda t}{p_{0}}^3
\}) exp(\frac{i\lambda t}{\hbar}\{q_{3}{\pi_q}_{3}-p_{3}{\pi_p}_{3}\})\chi\nonumber\\
&=& exp(\frac{i\lambda}{2\hbar}\{e^{2\lambda
t}{v_{3}}^5+e^{-2\lambda t}{p_{3}}^9 \})\psi(e^{\lambda
 t}q_{3})\phi^\star(e^{-\lambda
 t}p_{3})e^{-\frac{i}{\hbar}p_{3}q_{3}}.
\end{eqnarray}
where , are unitary transformation of two consecztive canonical
transformation. Finally, eigenfunction of Eq. (15) reads
The above
eigenfunction are completed to investigate the uncertainties
relations for the combined systems actual and images oscillators
system.
\section{Uncertainty relations for actual and  image oscillators }
In this section we calculate the uncertainties in position and
momentum for the actual  and the image oscillators. We calculate
 the extended propagator [9] for the combined actual and the image
 oscillators as follows
\begin{eqnarray}
K(q,p,t,q_i,p_i,t_i)&=&(\frac{1}{9\pi i\hbar}
)[\frac{\omega'}{sin\omega'(t-t_i)}
]\nonumber\\
&\times& exp[\frac{1}{2} (\frac{\omega'
e^{\lambda(t+t_i)}}{sin\omega'(t-t_i)}
)\times\{e^{\lambda(t-t_i)}q^2(cos\omega'(t-t_i)
-\frac{\lambda}{\omega'}sin(\omega'(t-t_i))\nonumber\\
&+&e^{-\lambda(t-t_i)}q_i^2(cos\omega'(t-t_i)
+\frac{\lambda}{\omega'}sin(\omega'(t-t_i))-2qq_i\}]
\nonumber\\
&\times& exp[\frac{1}{2} (\frac{\omega'
e^{-\lambda(t+t_i)}}{sin\omega'(t-t_i)}
)\times\{e^{-\lambda(t-t_i)}p^2(cos\omega'(t-t_i)
+\frac{\lambda}{\omega'}sin(\omega'(t-t_i))\nonumber\\
&+&e^{\lambda(t-t_i)}p_i^2(cos\omega'(t-t_i)
-\frac{\lambda}{\omega'}sin(\omega'(t-t_i))-2pp_i\}].\label{11}
\end{eqnarray}
When $\lambda\to 0$ then Eq. (\ref{11}) reduces to the familiar
form of the undamped extended harmonic oscillator propagtor [8].
 We assume that the initial state function for combine system in ground state is
 $\chi_{00}(q,p,0)=(\pi\delta^2)^{-\frac{1}{2}}exp(-\frac{q^2+p^2}{2\delta^2})$
where $delta$ is the width of the extended wave packet. Then one
gets using Eq. ()
\begin{eqnarray}
\chi_{00}(q,p,t)&=&\int\int dq_idp_iK(q,p,t,q_i,p_i,0)\chi_{00}(q_i,p_i,0)\nonumber\\
&=&(\frac{\pi}{\delta^2})[\frac{1}{2\delta^2}-\frac{i}{2}
\frac{\omega'}{\hbar} (\frac{cos\omega' t}{sin\omega'
t}+\frac{\lambda}{\omega'})]^{-\frac{1}{2}}\nonumber\\&\times&
(\frac{\omega' e^{\l t}}{2\pi i\hbar sin\omega'
t})^{\frac{1}{2}}exp[-\frac{q^2}{2}
\{\frac{1}{\delta^2}e^{2\lambda
t}(1+[\frac{1}{\delta^4}(\frac{\hbar}{\omega'})^2+
2(\frac{\lambda}{\omega'} )^2-1]\nonumber\\&sin^2&\omega'
+\frac{\lambda}{\omega'}sin2\omega'
t)^{-1}-i\{\frac{\omega'}{\hbar} \frac{e^{2\lambda t}}{sin\omega'
t}\times([cos\omega' t-\frac{\lambda}{\omega'}sin\omega' t-
\nonumber\\&(&cos\omega' t+\frac{\lambda}{\omega'}sin\omega'
t)][1+\frac{1}{\delta^4}(\frac{\hbar}
{\omega'})^{2}+2(\frac{\lambda}{\omega'})^2-1]sin2\omega'
t+\frac{\lambda}{\omega'}sin2\omega' t \}\}^{-1}]
[\frac{1}{2\delta^2}+\frac{i}{2} \frac{\omega'}{\hbar}
(\frac{cos\omega' t}{sin\omega'
t}+\frac{-\lambda}{\omega'})]^{-\frac{1}{2}}\nonumber\\&\times&
(\frac{-\omega' e^{-\lambda t}}{2\pi i\hbar sin\omega'
t})^{\frac{1}{2}}exp[-\frac{p^2}{2}
\{\frac{1}{\delta^2}e^{-2\lambda
t}(1+[\frac{1}{\delta^4}(\frac{\hbar}{\omega'})^2+
2(\frac{\lambda}{\omega'} )^2-1]\nonumber\\&sin^2&\omega'
+\frac{-\lambda}{\omega'}sin2\omega'
t)^{-1}+i\{\frac{\omega'}{\hbar} \frac{e^{-2\lambda t}}{sin\omega'
t}\times([cos\omega' t+\frac{\lambda}{\omega'}sin\omega' t-
\nonumber\\&(&cos\omega' t-\frac{\lambda}{\omega'}sin\omega'
t)][1+\frac{1}{\delta^4}(\frac{\hbar}
{\omega'})^{2}+2(\frac{\lambda}{\omega'})^2-1]sin2\omega'
t-\frac{\lambda}{\omega'}sin2\omega' t
\}\}^{-1}]e^{-\frac{ipq}{\hbar}}.
\end{eqnarray}
Using Eq. (\ref{1}) the uncertainties of positions and momenta can
be computed for the actual  and the image oscillators as follows
\begin{eqnarray}
 <\Delta q>&=&\sqrt{<q^2>-<q>^2}\nonumber\\&=&\frac{\delta}{\sqrt{2}}e^{-\lambda t}\{1+[(
\frac{\sqrt{\hbar}}{\delta})^4
(\frac{1}{\omega'})^2+(\frac{\lambda}{\omega'})^2-1]sin^2\omega'
t+\frac{\lambda}{\omega'}sin2\omega' t\}^{\frac{1}{2}}.
\end{eqnarray}
and
\begin{eqnarray}
<\Delta\pi_q>
&=&\sqrt{<\pi_q^2>-<\pi_q>^2}\nonumber\\&=&\frac{\delta}{\sqrt{2}}e^{-\lambda
t}\{1+[( \frac{\sqrt{\hbar}}{\delta})^4
(\frac{1}{\omega'})^2+(\frac{\lambda}{\omega'})^2-1]sin^2\omega'
t-\frac{\lambda}{\omega'}sin2\omega' t\}^{\frac{1}{2}}.
\end{eqnarray}
and
\begin{eqnarray}
 <\Delta p>&=&\sqrt{<p^2>-<p>^2}\nonumber\\&=&\frac{\delta}{\sqrt{2}}e^{\lambda t}\{1+[(
\frac{\sqrt{\hbar}}{\delta})^4
(\frac{1}{\omega'})^2+(\frac{\lambda}{\omega'})^2-1]sin^2\omega'
t-\frac{\lambda}{\omega'}sin2\omega' t\}^{\frac{1}{2}}.
\end{eqnarray}
and
\begin{eqnarray}
<\Delta\pi_p>
&=&\sqrt{<\pi_p^2>-<\pi_p>^2}\nonumber\\&=&\frac{\delta}{\sqrt{2}}e^{\lambda
t}\{1+[( \frac{\sqrt{\hbar}}{\delta})^4
(\frac{1}{\omega'})^2+(\frac{\lambda}{\omega'})^2-1]sin^2\omega'
t+\frac{\lambda}{\omega'}sin2\omega' t\}^{\frac{1}{2}},
\end{eqnarray}
The above results for actual and image oscillator, in separate
form, is in agreement with those obtained by Bateman. However,
when $\lambda\neq0$, it is not possible to separate the
oscillators as Bateman  does and the Heisenberg uncertainty
relations would not hold for each oscillator, separately, (see
figs. 1 and 2). In fact the uncertaintyrelations would also looked
upon for combined system n extended phase space as shown in fig.
3.
 \clearpage
\begin{figure}
\vskip2truecm \hskip1.truecm \epsfxsize=8.5truecm \epsffile{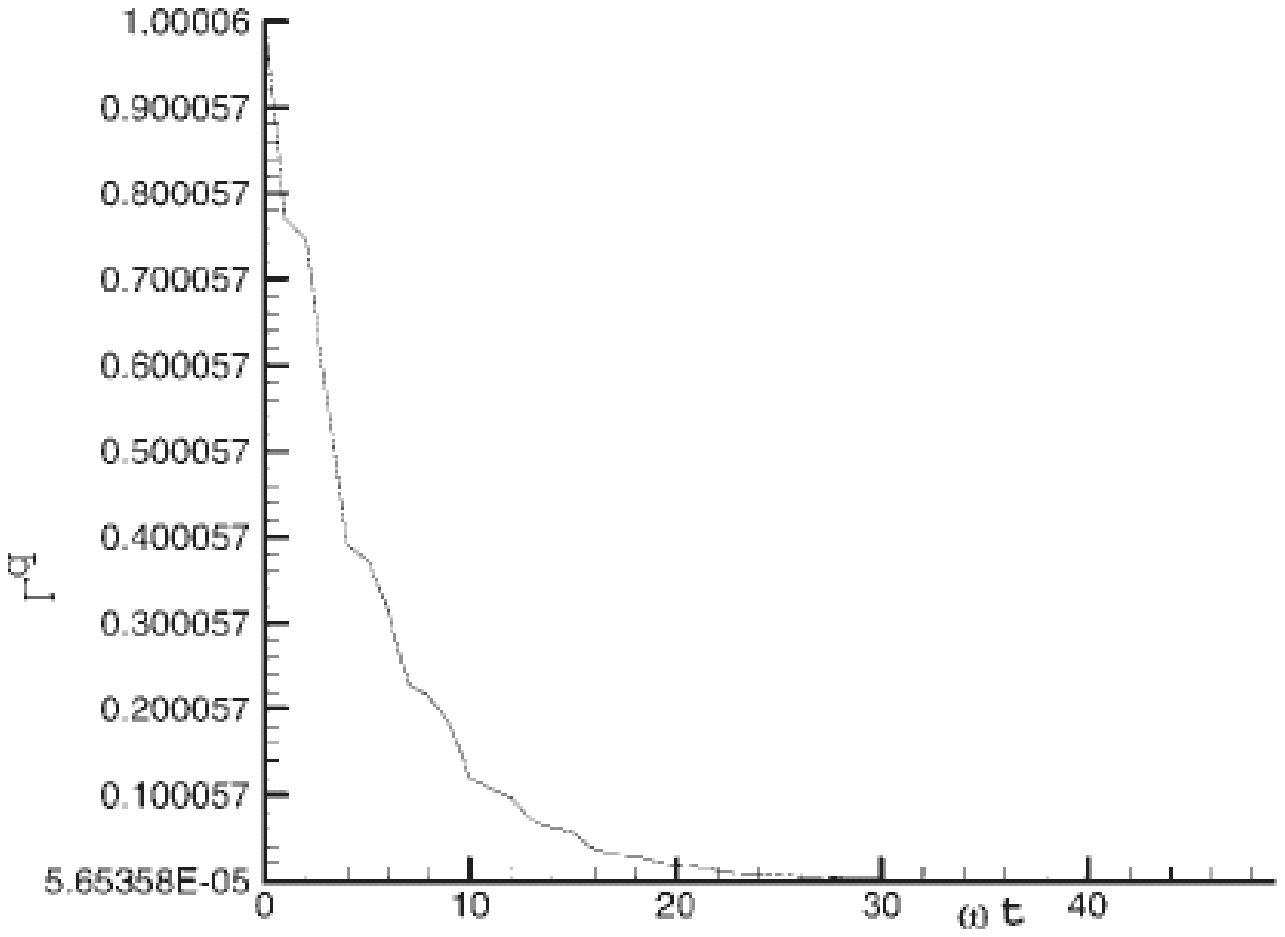}
\hskip0.6truecm \epsfxsize=7.5truecm\epsffile{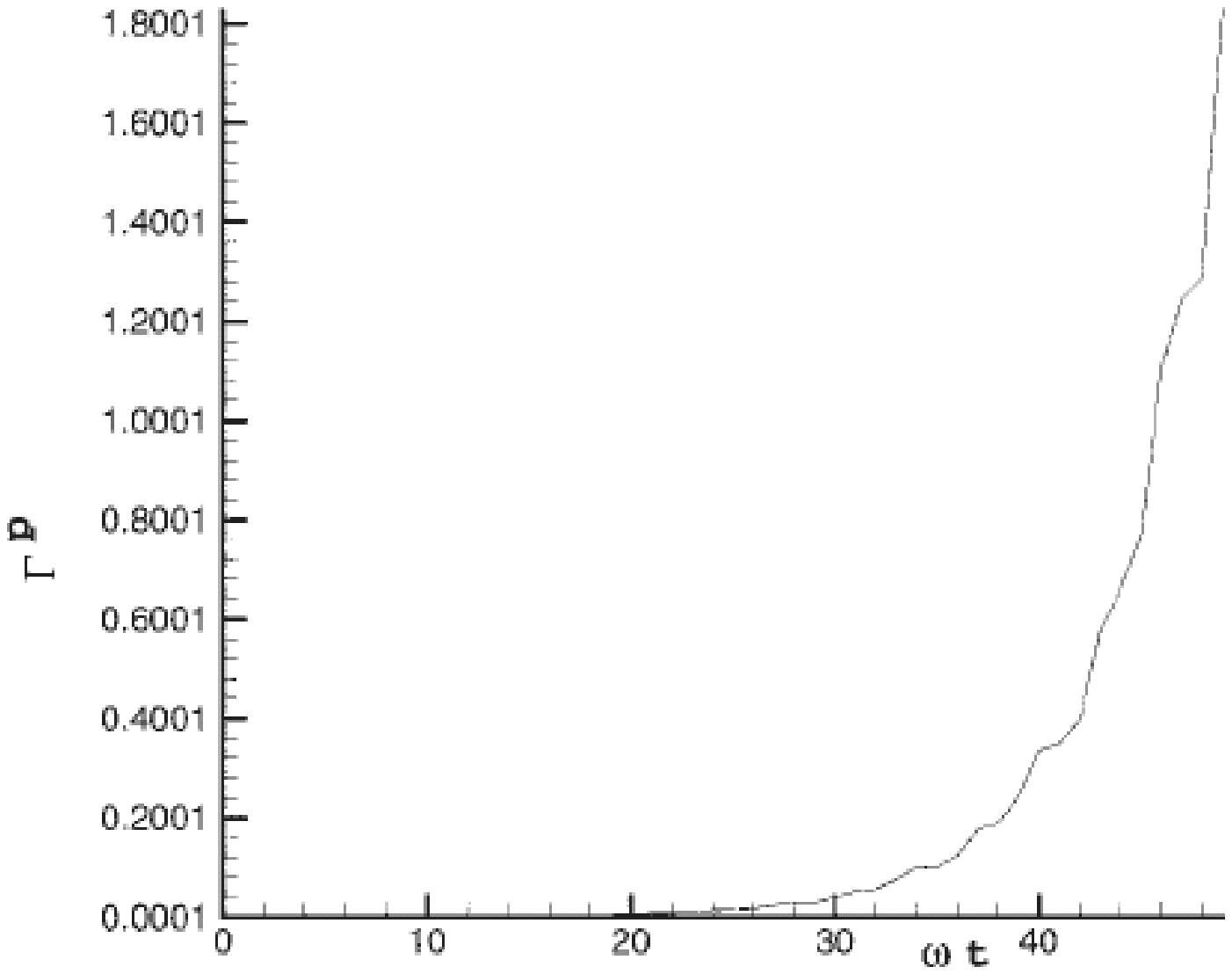} 
\centerline{\epsfxsize=7.5truecm \epsffile{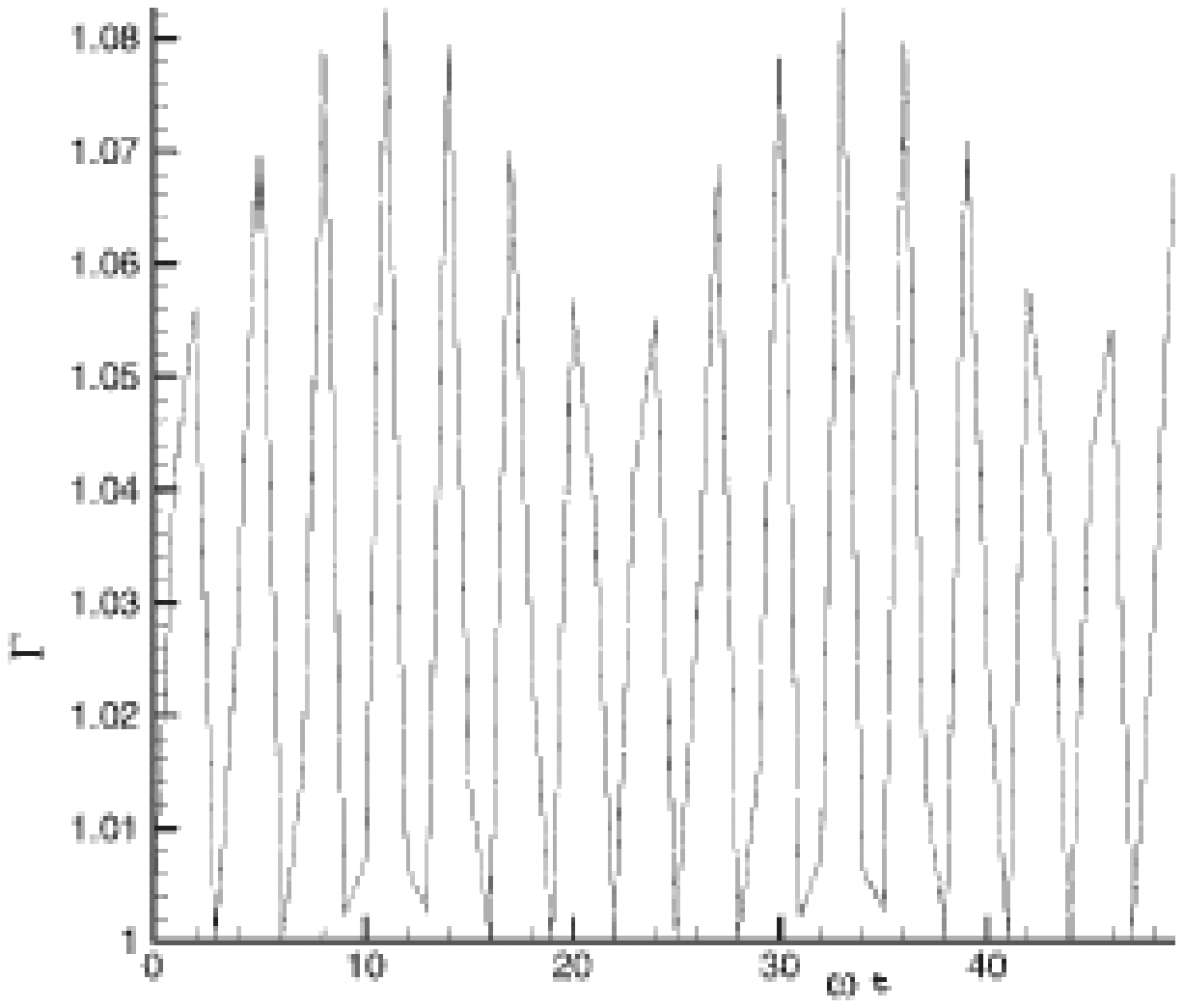}} 
\caption{Uncertainty relation for a) actual oscillator, b) image
oscillator and c) for combined system (actual and image
oscillator) }
\end{figure}

\LastPageEnding

\end{document}